\documentclass[twocolumn,showpacs,amsmath,amssymb,prl]{revtex4-1}
\usepackage{graphicx,amsmath,amssymb,bm,color,engord,soul}
\usepackage[usenames,dvipsnames]{xcolor}
\usepackage[colorlinks,colorlinks,urlcolor=Magenta,citecolor=Magenta,linkcolor=black]{hyperref}

\begin{document}
\title{Ultrafast quantum random access memory utilizing single Rydberg atoms in a 
Bose-Einstein condensate}
\author{Kelly R. Patton and Uwe R. Fischer}
\affiliation{Seoul National University, Department of Physics and Astronomy\\ Center for Theoretical Physics, 151-747 Seoul, Korea }
\date{\today}
\begin{abstract}
We propose a long-lived and rapidly accessible quantum memory unit, for which the operational Hilbert space is spanned by states involving the two macroscopically occupied hyperfine levels of a miscible binary atomic Bose-Einstein condensate  
and the Rydberg state of a single atom. 
It is shown that an arbitrary qubit state, initially prepared using a flux qubit, can be rapidly transferred to and from the trapped atomic ensemble in approximately 10 ns and with a large fidelity of 97\%, via an effective two-photon process using an external laser for the transition to the Rydberg level.
The achievable ultrafast transfer of quantum information therefore enables a large number 
of storage and retrieval cycles from the highly controllable quantum optics setup
of a dilute ultracold gas, even within the typically very short flux qubit lifetimes of the order of 
microseconds.
\end{abstract}
\maketitle

The intense cross-disciplinary interest in the field of quantum information \cite{Nielsen00}  has  been mainly driven by the promise of using quantum computers to solve compelling  mathematical problems, such as prime factorization \cite{Shor97},  exponentially faster than currently known classical algorithms.  Subsequently, this has led to the rapid development of new quantum algorithms, as well as the  search for a physical qubit architecture processing fast logic gates, rapid logical operations being necessary to manipulate  quantum information before coherence is lost. 

The experimental realization of quantum computers often involves developing auxiliary functionality, such as memory qubits and buses \cite{Mariantoni11}, in  addition to the implementation of error correction, logic gates, and measurement processes.  One such unit, the quantum analog of random access memory (qRAM) can be used to coherently store quantum information, read the stored information by measurements, and possibly even erase, or zero out, the stored data. 
Hybrid qubit architectures, where qubits are realized by physically distinct systems, which are allowed to interact,  have been some of the most interesting and promising candidates for qRAMs.  Various examples for such hybrid systems include cavity-QED with Bose-Einstein condensates (BECs) \cite{Brennecke1,Brennecke2}, nanomechanical resonators coupled to a superconducting phase qubit \cite{OConnell,Cleland}, solid state nuclear spins interacting with flux qubits \cite{Marcos,Zhu,JulsgaardPRL13},  and ultracold atoms coupled to a superconducting waveguide cavity \cite{Verdu}.

Recently a hybrid producing a high fidelity quantum memory unit has been suggested \cite{PattonPRA13,PattonEPL13}, which involves the long-lived \cite{Bernon13} hyperfine states of an  atomic BEC.   It was shown theoretically that one could transfer and store  an arbitrary qubit in the hyperfine states that was initially prepared  in a magnetically coupled flux qubit using a superconducting quantum interference device (SQUID).  The BEC-SQUID coupling is generated by the coherent macroscopic electromagnetic fields produced by the circulating current states of the SQUID. These currents can be used to induce magnetic dipole transitions between  the hyperfine split ground states of a trapped ultracold gas.  The presence of the $N$-atom BEC provides a bosonic enhancement of the single-atom magnetic dipole  Rabi frequency $\Omega_{\rm single}$ between the two hyperfine levels. This leads to a BEC-SQUID  Rabi frequency of  $\Omega_{\rm BS}=\sqrt{N}\Omega_{\rm single}$.  For typical atomic  BEC densities this can increase the single-atom Rabi frequency by a factor of $10^{3}$. 
The time to transfer a qubit state from the SQUID to the BEC is approximately half a Rabi cycle $\tau=\frac{\pi}{2}\Omega^{-1}_{\rm BS}$, leading to 
$\tau\approx 1.5\, \mu{\rm s}$ \cite{PattonPRA13}. 
This transfer time is of the same order as the coherence times of current flux qubits.   Thus, for such a system to be a useful element of a functional quantum computer either the BEC-SQUID coupling would have to be increased, the coherence times of flux qubits improved, or some combination of both by several orders of magnitude.

\begin{figure}[b]
\vspace{0.5em}
\includegraphics[width=0.95\columnwidth]{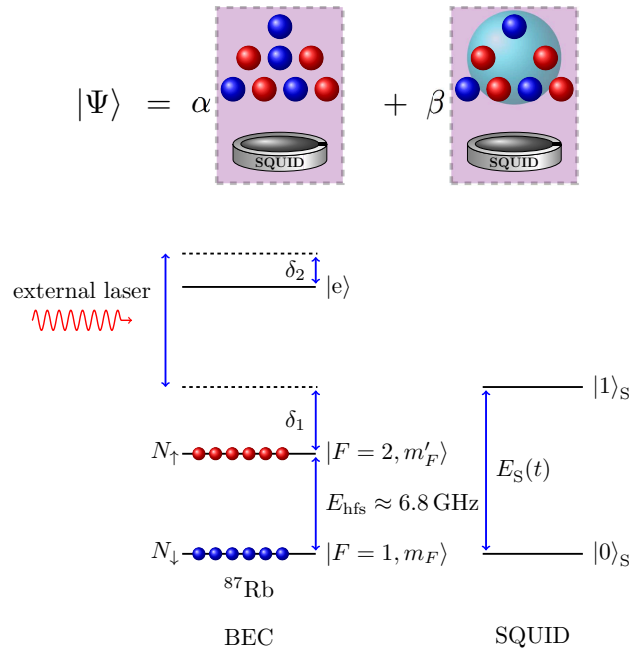}
\caption{Top: A general qubit state of the  BEC-Rydberg system involves the absence or presence of a single Rydberg atom in the BEC. 
Bottom: The energy level schematic for the proposed  BEC-Rydberg qubit coupled to a SQUID and an external laser source. When the SQUID's energy is dynamically adjusted  such that 
$E_{\rm S}(t)=\omega_{\uparrow}-\omega_{\downarrow}+\delta_1=E_{\rm hfs}+\delta_1$, this effectively 
induces, by de-excitation of the SQUID and the optical transition caused by the external laser, two-photon Rabi oscillations connecting the  state  $|F=1,m_{F}\rangle$ of rubidium-87, and a suitably chosen long-lived Rydberg state $|\rm e\rangle$.  The presence of the second macroscopically occupied hyperfine state $|F=2,m'_{F}\rangle$ further enhances the BEC-SQUID single-particle Rabi frequency between the two hyperfine levels, yielding the ${\cal O}(N)$ expression 
$\Omega_{\rm BS}\simeq\sqrt{N_{\uparrow}N_{\downarrow}}\Omega^{\rm BS}_{\rm single}$. \label{fig1}}
\vspace*{-1em}
\end{figure}

Here, we propose a 
memory qubit architecture involving an ultracold trapped atomic gas coupled to a flux qubit, which overcomes the aforementioned limitations; it can process both relatively  
long coherence times for the quantum information, stored in Rydberg states of atoms in a BEC,    
and ultrafast state transfer times of order tens of nanoseconds. 
This is accomplished in part by trapping a miscible binary BEC, for example using the two hyperfine  states $|\downarrow\rangle = |5^{2}S_{1/2},F=1,m^{}_{F}=-1\rangle$ and $|\uparrow\rangle =|5^{2}S_{1/2},F=2,m^{}_{F}=-2\rangle$ of $^{87}\!$Rb \cite{Hoefer}.   Then if each hyperfine state is populated with $N_{\uparrow}$ and $N_{\downarrow}$ atoms respectively, the Rabi frequency between these two states, induced by coupling to the SQUID, scales as $\Omega_{\rm BS}\simeq\sqrt{N_{\uparrow}}\sqrt{N_{\downarrow}}\Omega_{\rm single}$. For $N_{\uparrow}=N_{\downarrow}=N/2$, $\Omega_{\rm BS}=(N/2)\Omega_{\rm single}$ \cite{numberimbalancefootnote}.  Although  this  alone increases the qubit transfer time, from the SQUID to the BEC, by a factor of $\sqrt{N}$ over the single component BEC, the coupled binary BEC and SQUID system no longer spans a simple two-qubit Hilbert space.  For example, if the qubit states of the BEC are taken to be $|0\rangle_{\rm B}=|N_{\uparrow},N_{\downarrow}\rangle$ and  $|1\rangle_{\rm B}=|N_{\uparrow}+1,N_{\downarrow}-1\rangle$, then the  ``ground state'' of the coupled system $|00\rangle=|0\rangle_{\rm B}\otimes|0\rangle_{\rm S}$ can easily transition to  $|N_{\uparrow}-1,N_{\downarrow}+1\rangle\otimes|1\rangle_{\rm S}$ (when on resonance), effectively leaving the computational two-qubit Hilbert space.  To circumvent this while still maintaining  the enhancement of the BEC-SQUID Rabi frequency, we further couple the $\uparrow$-hyperfine state to a suitably chosen Rydberg state $|{\sf e}\rangle$ by an external laser source, see Fig.~\ref{fig1} for an energy level diagram. 
In the following we show this results in a two-photon Rabi transfer connecting the $\downarrow$-hyperfine state and the Rydberg state, while the filled $\uparrow$-states only remain to enhance the BEC-SQUID coupling, i.e., they are no longer part of the computational qubit basis.
The corresponding so-called {\em many-body Rabi} oscillations have been recently observed \cite{DudinNatPhys12}.   The manipulation of quantum information using Rydberg atoms is one of the major active schemes toward  realizing a quantum computer \cite{SaffmanRMP}, as well as the development of solid-state superconducting qubits \cite{DevoretScience}.  The proposed qRAM hybridizes these two well-known setups. 
We note that while two-photon mediated architectures for quantum information processing involving a charge qubit, an electromagnetic resonator, and  a trapped gas of polar molecules have also been proposed in \cite{RablPRL2006,TordrupPRL2008}, they are much more demanding as regards their experimental implementation. In particular, the necessity of trapping ultracold polar molecules very near ($\sim 10\,\mu$m) to a relatively hot and electronically active waveguide surface is a serious obstacle. For the present architecture, the corresponding requirements for trapping neutral atoms are much less severe and, in addition, the experimental procedures are well established, see below for a more detailed discussion.

Neglecting atom-atom interactions the second quantized Hamiltonian for a trapped atomic ultracold  gas is taken to be $(\hbar=1)$
\begin{align}
\label{eq: full BEC Hamiltonian}
\hat{H}_{\rm B}&=\sum_{\sigma}\int d{\bm r}\, \hat{\Psi}^{\dagger}_{\sigma}({\bm r})\bigg[-\frac{\nabla^{2}}{2m}+\omega_{\sigma}+V_{\rm trap}({\bm r})\bigg]\hat{\Psi}^{}_{\sigma}({\bm r})\nonumber\\&+\int d{\bm r}\,\hat{\Psi}^{\dagger}_{\rm e}({\bm r})\bigg[-\frac{\nabla^{2}}{2m}+\omega_{\rm e}+V_{\rm trap}({\bm r})\bigg]\hat{\Psi}^{}_{\rm e}({\bm r}),
\end{align}
where $\omega_{i}$ labels the internal energy of each species, e.g. the combined hyperfine and Zeeman energy of each spin or the Rydberg state energy.  The neglect of interactions among the two BEC spin states assumes the two components are miscible and all interaction renormalized energies are small compared to the hyperfine energy level spacings. 
The atom-atom interaction effects of a single Rydberg atom immersed in a BEC sea are an emerging field of study.  
For example, a recent experimental study investigated highly excited Rydberg electrons in a BEC, interacting essentially with the 
condensate cloud \cite{Balewski}, and Ref.~\cite{MiddelkampPRA07} calculates the  mean-field energy shifts of the Rydberg states due to the condensate. 
These intrinsic effects, and interaction induced decoherence, are however probably much smaller than those induced by the proximity of the system to the surface on which the SQUID resides.  These surface effects \cite{FortaghRMP,HohenesterPRA,ZhangEPJD} will be discussed below.  The Hamiltonian of the flux qubit is described by the effective low-energy dynamics of a SQUID. In the $|{\sf L}\rangle$ and $|{\rm R}\rangle$ current basis \cite{SQUIDbook}
\begin{equation}
\label{eq: flux qubit Hamiltonian in current basis}
\hat{H}_{\rm S}(t)=\frac{\varepsilon}{2}\sigma_{z}-\frac{\Delta(t)}{2}\sigma_{x},
\end{equation}
or in its energy eigenbasis $|0\rangle_{\rm S}$ and $|1\rangle_{\rm S}$
\begin{equation}
\label{eq: flux qubit Hamiltonian in eigen basis}
\hat{H}_{\rm S}(t)=\frac{E_{\rm S}(t)}{2}\sigma_{z}.
 \end{equation}
 The time dependence of the tunneling amplitude $\Delta(t)$ allows for the dynamic control of the energy level spacing.  This has been shown to be experimentally feasible on a subnanosecond time scale \cite{PaauwPRL, CastellanoNJP}.  Assuming all other transitions are far off resonance, the interaction between the two BEC hyperfine states and the SQUID is described by a single magnetic dipole coupling term 
 \begin{equation}
 \label{eq: BS coupling}
 \hat{H}_{\rm BS}=-\sum_{\sigma,\sigma'}\int d{\bm r}\,\hat{\Psi}^{\dagger}_{\sigma}({\bm r})\boldsymbol{\mu}_{\sigma,\sigma'}\hat{\Psi}^{}_{\sigma'}({\bm r})\otimes \hat{\bm B}({\bm r}),
 \end{equation}
 where $\boldsymbol{\mu}_{\sigma,\sigma'}$ is the total magnetic moment of an atom, and $\hat{\bm B}({\bm r})$ is the macroscopic  magnetic field operator of the SQUID.  The quantum mechanical nature of the magnetic field is inherited from the macroscopic quantum mechanical current carrying states, which are eigenstates of the current operator $\hat{I}\simeq I|{\sf L}\rangle\langle {\sf L}|-I|{\rm R}\rangle\langle {\rm R}|=I\sigma_{z}$.  If ${\bm B}(\bm r)$ is the classical magnetic field of a current carrying loop, which models the SQUID, then $\hat{\bm B}({\bm r})={\bm B}(\bm r)\sigma_{z}$, see Ref.~\cite{PattonPRA13} for details.   The coupling of the upper hyperfine state to the Rydberg level is done via an external laser source.  Within the electric dipole approximation 
 \begin{equation}
 \label{eq: external coupling}
 \hat{H}_{\rm ex}(t)=-\sum_{i,j}\int d{\bm r}\, \hat{\Psi}^{\dagger}_{i}({\bm r}){\bm d}_{i,j}\hat{\Psi}^{}_{j}({\bm r})\cdot {\bm E}_{\rm ex}({\bm r},t)\otimes \openone.
 \end{equation}
 Here, ${\bm d}_{i,j}$ are the electric dipole moment matrix elements connecting the two states $i,j \in \{\uparrow, {\rm e}\}$.  The plane wave electric field of a laser polarized in the $z$-direction can be written as
$ {\bm E}_{\rm ex}({\bm r},t)=E_{\rm ex}{\bm e}_{z}e^{i({\bm k}\cdot{\bm r}-\omega t)}+{\rm c.c.}
\equiv {\bm E}^{(+)}_{\rm ex}({\bm r})e^{-i\omega t}+{\bm E}^{(-)}_{\rm ex}({\bm r})e^{i\omega t}.$  
The frequency of the light is fixed to $\omega=\delta_{2}-\delta_{1}+\omega_{\rm e}-\omega_{\uparrow}$, see Fig.~\ref{fig1}.
 The total Hamiltonian is then 
 \begin{equation}
 \label{eq: total Hamiltonian}
 \hat{H}(t)=\hat{H}_{\rm B}\oplus \hat{H}_{\rm S}(t)+\hat{H}_{\rm BS}+\hat{H}_{\rm ex}(t). 
 \end{equation}
\pagebreak
 This Hamiltonian  can be represented as a matrix in the occupation basis $|N_{\downarrow},N_{\uparrow},N_{\rm e}\rangle_{\rm B}\otimes |i\rangle_{\rm S}$. 
Because the BEC-SQUID coupling term  Eq.\,\eqref{eq: BS coupling} only connects states that differ by single pseudo-spin and flux qubit quanta, at zero temperature the full Hilbert space can be approximated by a six-dimensional  subspace  spanned by 
\begin{equation}
\label{eq: Hilbert space subspace}
\begin{split}
|11\rangle&=|N_{\downarrow}-1,N_{\uparrow},1\rangle^{}_{\rm B}\otimes |1\rangle^{}_{\rm S},\\
|10\rangle&=|N_{\downarrow}-1,N_{\uparrow},1\rangle^{}_{\rm B}\otimes |0\rangle^{}_{\rm S},\\
|{\rm v}1\rangle&=|N_{\downarrow}-1,N_{\uparrow}+1,0\rangle^{}_{\rm B}\otimes |1\rangle^{}_{\rm S},\\
|{\rm v}0\rangle&=|N_{\downarrow}-1,N_{\uparrow}+1,0\rangle^{}_{\rm B}\otimes |0\rangle^{}_{\rm S},\\
|01\rangle&=|N_{\downarrow},N_{\uparrow},0\rangle^{}_{\rm B}\otimes |1\rangle^{}_{\rm S},\\
|00\rangle&=|N_{\downarrow},N_{\uparrow},0\rangle^{}_{\rm B}\otimes |0\rangle^{}_{\rm S}.
\end{split} 
\end{equation}
This restriction  of the Hilbert space  neglects possible transitions such as $|N_{\downarrow},N_{\uparrow},N_{\rm e}\rangle_{\rm B}\otimes|0\rangle_{\rm S}\to|N_{\downarrow}+1,N_{\uparrow}-1,N_{\rm e}\rangle_{\rm B}\otimes|1\rangle_{\rm S}$ and $|N_{\downarrow},N_{\uparrow},N_{\rm e}\rangle_{\rm B}\otimes|0\rangle_{\rm S}\to|N_{\downarrow},N_{\uparrow}-1,N_{\rm e}+1\rangle_{\rm B}\otimes|0\rangle_{\rm S}$.  This is valid only if the energy spacing of the flux qubit is far off resonance from the hyperfine splitting for all times, and the detuning $\delta_{1}$ is large, see Fig.~\ref{fig1}.   The index v in \eqref{eq: Hilbert space subspace} labels the virtual  intermediate state of the atomic system, while the remaining states $|00\rangle, |01\rangle,|10\rangle$, and $|11\rangle$ span the qubit-qubit Hilbert space.  In other words, the  ultracold atomic gas qubit is given by the two component BEC ground state $|0\rangle_{\rm B}$ and a system with the binary BEC and a single Rydberg atom $|1\rangle_{\rm B}$.

The Zeeman energy shifts from the BEC-SQUID coupling \eqref{eq: BS coupling} simply lead to a small renormalization of the flux qubit energies and can be absorbed into $\varepsilon$.  
We assume in what follows,  for simplicity,  that the effective left-right current states of the SQUID are degenerate, i.e., we set $\varepsilon=0$ in Eq.\,\eqref{eq: flux qubit Hamiltonian in current basis}. 
For weak SQUID--BEC coupling, i.e., within the rotating wave approximation leading to the truncated basis Eq.\,\eqref{eq: Hilbert space subspace}, the Hamiltonian \eqref{eq: total Hamiltonian} can then be expressed as
\begin{widetext}
\begin{align}
\label{eq: subspace full Hamiltonian}
&\hat{H}(t)\simeq \left(\begin{array}{ccc}\omega_{\rm e}-\omega_{\downarrow} & -(\Omega_{\rm ex}e^{-i\omega t}+\Omega_{\rm ex}e^{i\omega t})& 0 \\-(\Omega^{*}_{\rm ex}e^{i\omega t}+\Omega^{*}_{\rm ex}e^{-i\omega t})  & \omega_{\uparrow}-\omega_{\downarrow}  & 0 \\ 0 & 0 & 0 \end{array}\right)\oplus \frac{\Delta(t)}{2}\sigma_{z}-\left(\begin{array}{ccc}0 & 0 & 0 \\0 & 0 & \Omega_{\rm BS} \\0 & \Omega^{*}_{\rm BS} & 0\end{array}\right)\otimes\sigma_{x}.
\end{align}
\end{widetext}
The complex single-photon Rabi frequencies are given by  the expressions 
$\Omega_{\rm ex}\simeq \sqrt{N_{\uparrow}}\sum_{i,j}\int d{\bm r}\phi^{*}_{i}({\bm r}){\bm d}_{i,j}\phi^{}_{j}({\bm r})\cdot {\bm E}^{(+)}({\bm r})\equiv \sqrt{N_{\uparrow}}\Omega^{\rm ex}_{\rm single}$, $\Omega_{\rm BS}\simeq  \sqrt{N_{\uparrow}} \sqrt{N_{\downarrow}}\sum_{\sigma,\sigma'}\int d{\bm r}\phi^{*}_{\sigma}({\bm r}){\boldsymbol \mu}_{\sigma,\sigma'}\phi^{}_{\sigma'}({\bm r})\cdot {\bm B}^{}({\bm r})\equiv \sqrt{N_{\uparrow}} \sqrt{N_{\downarrow}}\Omega^{\rm BS}_{\rm single}$, where $\phi_{\alpha}({\bm r})$ are the atomic center of mass wave functions of the trapped gas for each species.   Furthermore to arrive at \eqref{eq: subspace full Hamiltonian} overall constants and other subdominant Rabi couplings on the order of a single-particle term have been neglected. 

For the Hamiltonian \eqref{eq: subspace full Hamiltonian} and within the rotating wave approximation one obtains an effective two-photon Rabi process, cf. Fig.\,\ref{fig1}. Connecting the 
 states $|01\rangle$ and $|10\rangle$ with a minimal occupation of other levels can be achieved by setting $\delta_{2}=\delta^{-1}_{1}(|\Omega_{\rm BS}|^{2}-|\Omega_{\rm ex}|)$ and $\delta_{1}\gg |\Omega_{\rm ex}|$,  $|\Omega_{\rm BS}|$ \cite{GentilePRA}. The two-photon Rabi frequency is then given by $\Omega=|\Omega_{\rm ex}||\Omega_{\rm BS}|/\delta_{1}.$ 
An estimate of the two-photon transfer time $\tau=\frac{\pi}{2}\Omega^{-1}$ can now be found using experimental parameters as follows,   $N_{\uparrow}=N_{\downarrow}=N/2\sim 10^{6}$, $\Omega^{\rm ex}_{\rm single}\sim 1.0$ MHz, and, for a SQUID loop  carrying a $1\, \mu $A current with a radius of $1\, \mu$m and  a  BEC-SQUID separation of 25 $\mu \rm m$, $\Omega^{\rm BS}_{\rm single}\sim 1.0$ kHz.  Setting $\delta_{1}=10\,\Omega_{\rm BS}$ gives $\tau\sim 10$ ns.  This is a reduction of the transfer time by a factor of roughly $10^{3}$ over the single BEC system proposed in Ref.~\cite{PattonPRA13}.   As can be seen from Fig.~\ref{fig1}, for the two systems to be on resonance requires $E_{\rm S}=E_{\rm hfs}+\delta_{1}$. For  $^{87}$Rb, $E_{\rm hfs}\approx 6.8$ GHz, and  $\delta_{1}\sim 10.0$ GHz; thus, a relatively large SQUID frequency of approximately $20$ GHz is required. Ref.\,\cite{PaauwPRL} reports that 14 GHz 
has already been achieved, so that this is within the range of current or near future technology. 

Next we calculate the fidelity of a state transfer from the flux qubit to the BEC-Rydberg qubit.  The state transfer is achieved by initially  preparing the SQUID in an arbitrary  coherent state $|\Psi_{0}\rangle=\alpha|00\rangle+\beta|01\rangle$.  The SQUID's energy level spacing is then dynamically adjusted, such that $E_{\rm S}(t)\simeq E_{\rm hfs}+\delta_{1}$ for half a Rabi cycle $\tau$ and then brought far off resonance again.  Within \eqref{eq: subspace full Hamiltonian}  this is done by setting $\Delta(t)=(E_{\rm hfs}+\delta_{1})W(t)$, where the function $W(t)$ smoothly ramps the two systems into and out of resonance \cite{Window function}.  
The time-dependent fidelity of a single state transfer is taken to be the overlap of the time evolved initial state $|\Psi_{0}\rangle$ with respect to the full Hamiltonian  and that of the time evolved target state $|\Psi_{\rm target}\rangle=\alpha|00\rangle+\beta|10\rangle$ under the uncoupled Hamiltonian $\hat{H}_{0}(t)=\hat{H}_{\rm B}\oplus\hat{H}_{\rm S}+\hat{H}_{\rm ex}(t)$;
\begin{equation}
\label{eq: single state fidelity}
F(t)=|\langle \Psi_{\rm target}(t)|\Psi(t)\rangle|,
\end{equation}
where $|\Psi(t)\rangle=\hat{U}_{H}(t)|\Psi_{0}\rangle$, and $|\Psi_{\rm target}(t)\rangle=\hat{U}_{H_{0}}(t)|\Psi_{\rm target}\rangle$. A general initial  state on the SQUID Bloch sphere can be written using polar coordinates as 
\begin{equation}
|\Psi_{0}\rangle=\cos(\theta/2)|00\rangle+e^{i\phi}\sin(\theta/2)|01\rangle,
\end{equation}
with $0\le \theta \le \pi$ and $0\le \phi \le 2\pi$ and similarly for the target state.  Averaging the fidelity over the Bloch sphere, 
\begin{equation}
\label{eq: averaged fidelity}
F_{\rm avg}(t)=\frac{1}{2\pi^{2}}\int\limits_{0}^{\pi}d{\theta}\int\limits_{0}^{2\pi}d\phi\, F(t).
\end{equation}
\begin{figure}
\includegraphics[width=0.95\columnwidth]{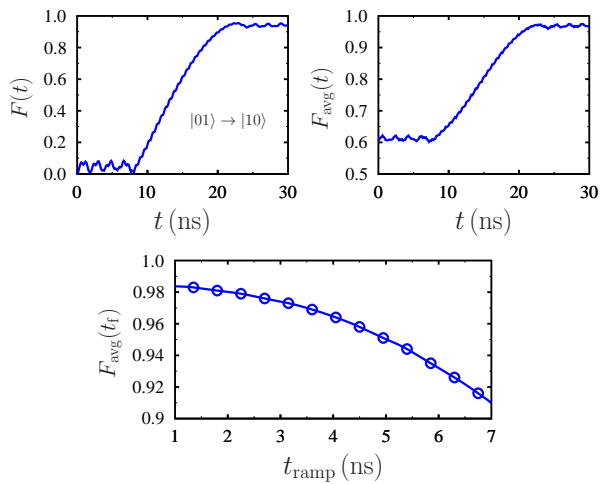}%
\caption{ For an atomic $^{87}$Rb system and  parameters listed in Tab.~\ref{tab: table1}, the top left plot shows the time-dependent fidelity \eqref{eq: single state fidelity} of the simple state transfer $|01\rangle\to |10\rangle$, and the  right shows the Bloch-sphere averaged fidelity \eqref{eq: averaged fidelity}. The final fidelities are $F(t_{\rm f})\approx 0.94$ and $F_{\rm avg}(t_{\rm f})\approx 0.97$. The final averaged fidelity as a function of ramp time is shown in the bottom panel. \label{fig2}}
\end{figure}
\begin{table}
\centering
\begin{tabular}{c|c|c}
\hline\hline
Parameter &   $\times(E_{\rm hfs})$& $^{87}$Rb (GHz)\\
\hline
$\omega_{\rm e}-\omega_{\downarrow}$ & 100.0 & 68.0\\
$\omega_{\uparrow}-\omega_{\downarrow}$ & 1.0& 6.8\\
$\Omega_{\rm BS}$ & 0.15& 1.0\\
$\Omega_{\rm ex}$ & 0.15& 1.0\\
$\delta_{1}$ & 1.5& 10.2\\
$\delta_{2}$ & 0.0& 0.0\\
\hline
\end{tabular}
\caption{The numerical values for the parameters that enter the effective Hamiltonian \eqref{eq: subspace full Hamiltonian} used to simulate a state transfer.  In general these are given in units of the hyperfine splitting $E_{\rm hfs}$ and as an explicit example in frequency units for a $^{87}$Rb gas.  For reasons of numerical stability we have only chosen a large energy difference separating the Rydberg state from the hyperfine levels,  instead of a realistic value.  The results are insensitive to this choice as long this value is significantly larger than the hyperfine splitting.\label{tab: table1}}
\vspace*{-0.5em}
\end{table}
The time-dependent fidelity of a simple state transfer $|01\rangle\to |10\rangle$ and the averaged fidelity, for a system with physical parameters listed in Tab.~\ref{tab: table1}, are shown in the top plots of Fig.~\ref{fig2}.  As one can see, within current experimental setups one can  obtain both a high fidelity and fast transfer using a two-photon coupling.  
The fidelity also depends on the time scale over which the two qubits are brought into and out of resonance.  This is called the ramp time $t_{\rm ramp}$.  Here, the ramp time is defined as the time it takes the flux qubit's energy spacing to change from being  1\% larger than its off resonance value to being within  99\%  of its on  resonance value.  The bottom panel of Fig.~\ref{fig2} shows how the final average fidelity changes as a function of the ramp time.  

We now come to discuss the experimental implementation of the present qRAM scheme.    
The radiative lifetimes, or qubit $T_{1}$ times, of (isolated) Rydberg atoms scales as the cube of the principal quantum number $n^{3}$ \cite{BrandenJPB10}.  For the proposed system, choosing a high Rydberg level has to take into account the increase of the orbital size of the state, which scales as $n^{2}$.  
For a rather moderate $n\sim 40$, $T_{1}\sim40\,\mu$s, which is already an order of magnitude longer than the phase coherence times of current flux qubits \textcolor{blue}{\cite{coherencenote}}.
Even though the BEC has to be in relative proximity of the SQUID, and its accompanying electronics,  
heating, decoherence, and atom loss due of the BEC are negligibly small at superconducting temperatures \cite{Henkel}.  

A potential complication, stemming from bringing the Rydberg atom close to a solid-state surface, is caused by the presence of electric fields originating from the adsorbates of atoms deposited on the surface, leading to nontrivial Stark shifts of the internal energy levels  of the trapped atoms \cite{TauschinskyPRA10,Hattermann}, and potentially also affect the phase coherence of the Rydberg state qRAM.  These shifts vary within the BEC cloud, depending on an atom's distance from the surface (Ref.\,\cite{Hattermann} quotes a shift per distance $\sim$ 1 MHz/$\mu$m at $n=35$).  Averaging over the size of the BEC, this effectively leads to  increased energy level ``line-widths''.   These line-widths could be larger than the band-width of the external laser source, which would diminish the bosonic enhancement of the single-particle Rabi frequency, i.e., the $\sqrt{N}$-dependence to some degree, as the number of atoms on resonance with the laser is reduced.  This effect can be minimized, though, by making the BEC highly oblate, 
with the weakly confining axes parallel to the solid state surface.   

In conclusion, we have proposed using a trapped atomic binary BEC with single Rydberg-excited atoms as a fast, long-lived, and functional qRAM for a flux qubit.  The presence of the two-component BEC  allows for a rapid high fidelity two-photon-mediated state to be transferred between the two systems, i.e., write the flux qubit to memory.  Reading  the memory qubit by quantum tomography can be done via photoionization of the Rydberg level and subsequent ion detection \cite{DudinNatPhys12,Stibor,CiampiniPRA}. This can also be used to erase the stored information. We have discussed that this setup appears feasible within present or near future technology.  

This research was supported by the NRF Korea, Grants No. 2010-0013103 and No. 2011-0029541. 


\begin{thebibliography}{99}
\bibitem{Window function} We choose $W(t)=\Theta(\kappa,t)\Theta(\kappa,\tau-t)+E^{}_{\rm or}\Theta(\kappa,-t)+E^{}_{\rm or}\Theta(\kappa,t-\tau)$, 
where $\Theta(\kappa,t)=\frac{1}{2}+\frac{1}{2}\tanh(\kappa t)$, $\tau$ is the time the two systems are on resonance,  
and  $E^{}_{\rm or}$ is the off-resonance value of 
$\Delta$ in units of $E_{\rm hfs}$ (we set $E^{}_{\rm or}=1/2+\delta_{1}/E_{\rm hfs}$. ) 
The function $\Theta(\kappa,t)$ is a smoothed step function, such that $\lim_{\kappa\rightarrow\infty}\Theta(\kappa,t)=\Theta(t)$.
\bibitem{coherencenote} Note, however, that the coherence lifetimes of Rydberg states in the condensate, depending on the 
latter's density, can be significantly lower than the radiative lifetimes, cf. \cite{Balewski}.
\end{thebibliography}
\end{document}